# The Giza "written" landscape and the double project of King Khufu

Giulio Magli,
Faculty of Civil Architecture,  Politecnico di Milano
P.le Leonardo da Vinci 32, 20133 Milano, Italy.
giulio.magli@polimi.it

*In recent years, different scientific disciplines - from Physics to Egyptology, from Geology to Archaeoastronomy - evidenced a series of clues pointing to the possibility that the original project of the pyramid complex of Khufu at Giza included also the project of the second pyramid complex at the same site, that of Khafra. The aim of the present paper is to discuss this issue in a systematic fashion and to explore its consequences on the way the ancient Egyptians conceived and built monuments and entire landscapes during the Old Kingdom.*

# 1. Introduction

The Giza plateau is perhaps the most studied and world-renewed archaeological site on earth. Children easily learn that the kings Cheops, Chepren and Mycerinus (Khufu, Khafra and Menkaura) subsequently built their pyramids there, and this is what is usually considered the established archaeological truth. However, it is not precisely so, and in recent years, a series of new evidences – arising from "hard science" disciplines such as Physics and Geology, as well as from Egyptology itself – came out pointing to a different conclusion, namely that the original project of the pyramid complex of Khufu at Giza included also the project of the nearby pyramid complex of Khafra.

It is important to stress from the very beginning that this theory does not negate that the latter was Khafra's tomb. The hypothesis is that, at the (probably premature) death of Djedefra (Khufu's first successor) the second complex, already devised by Khufu's architects, was taken on and finished as his own tomb by Khafra, to whom it is firmly - and in fact, correctly - attributed by Egyptology. Therefore, the hypothesis has no effect whatsoever on the standard, accepted chronology of these three Pharaohs of the 4$^{th}$ dynasty; it only implies that the second pyramid complex at Giza was planned and begun some 25 years before than usually believed. From the historical point of view, however, this proposal has relevant consequences, for it helps in identifying the ground breaking "solar" innovations introduced by Khufu in the identification of the Pharaoh as a living god, and to put them in the correct perspective with respect to the huge architectural project devised by his father Snefru at Dahshur.

The aim of the present paper is to present in a systematic fashion all the hints which point to this idea; some of these appeared scattered in a series of publications (Magli 2003,2005,2008,2010,2011); important others are due to my friend and colleague Juan Belmonte (see Shaltout, Belmonte and Fekri 2007a,b, Belmonte and Shaltout 2009, Magli and Belmonte 2012), yet others are present – but unmarked as such – in the works by the American Egyptologist Mark Lehner (Lehner 1985a,b,1999). Finally, a few are presented here for the first time.

I will assume in what follows that the reader has some familiarity with the monuments of the Giza plateau **(Fig. 1)**; accurate descriptions of them can anyhow be easily found (e.g. Lehner 1999, Verner 2002); for a comprehensive approach to the conception of the sacred space in the Old Kingdom see Magli (2013).

# 2. The clues to a double Khufu project

## *1 Geo-morphology of the Giza Plateau*

Let us imagine the Giza plateau when it was still free of any significant construction (Lehner 1985a, Reader 2001). The Plateau forms a huge, projecting rock terrace where Khufu's pyramid, from now on G1, will be constructed; slightly to the south, a smooth terrace slopes gently from the horizon towards the east. Now let us imagine a king who wants to build an object made out of some two millions blocks of stone on this plateau, together with two megalithic temples connected by an ascending causeway. Given the sheer insanity of the project in itself, what is "natural" to do is to extract the stone from the front of the smooth slope, reserving for the causeway a stone rib in the centre, and then build the pyramid at the upper end of the terrace, regularizing the level to fill the slope under the foundations.

The result of such a logical procedure is indeed, a thing visible today on the Giza plateau.

Yes, it is the *second* Giza complex G2.

G2 is called *second* because what happened according to standard reconstructions is that the first pyramid to be constructed was G1, which is built on the steep terrace (**Fig.2**). Of course, this difficult choice at least left the entire plateau to the south-west free to be quarried for the stones. In spite of such a freedom, the blocks were anyway quarried only to the side of the (still to come) causeway of the second pyramid, so that it was *later* possible for the builders of G2 to obtain the causeway (whose direction, as we shall see, was fixed by rigorous requirements) out of the pre-existing rock instead of being obliged to...construct it from huge blocks of stone. Curiously enough, in turn, this is exactly what the builders of G1 previously had to do, since from the G1 terrace the cliff down to the valley is much more steep.

Clearly, something sounds a bit illogical in all this. Logic would rather require G2 to be the *first* pyramidal complex to be built on the plateau, as I ventured to propose 10 years ago (Magli 2003). However, such an hypothesis contrasts with the archaeological records, which show beyond any doubt that G1 was built by Khufu and that G2 was the tomb of his son and second successor Khafra. Eliminating the impossible, what remains must be the truth, as Doyle's *Sherlock Holmes* was used to say, and therefore the unique possible solution remains, that *both* complexes were designed by Khufu's architects.

As we shall now see, a impressive variety of other clues point in this direction.

## 2 The Snefru project at Dahshur

The father of Khufu, Snefru, built two enormous pyramids at Dahshur, today usually called the Red and the Bent pyramid (**Fig. 3**). Each of the two has a small funerary temple; both have a causeway (that of the Red Pyramid is unexcavated) and the Bent pyramid has a Valley Temple; that of the Red Pyramid, if existing, has never been excavated. Only the south pyramid, the Bent one, has a annexed pyramid located to the south, along the mid line of the main one. The chamber of this pyramid is too small for a burial, so the latter is the first example of a "satellite", probably meant for a statue designated to house the Ka of the deceased.

The Bent Pyramid is 189 metres wide base and 105 metres tall. The slope changes at 49 metres, with the initial inclination dropping off more than 10°. The north pyramid or Red Pyramid is a little larger at the base compared to her sister to the south, but the height is virtually identical due to the fact that the chosen slope was lower and equal to that of the upper part of the south one, a first hint as to the wish of the builders to transmit a unified message. The project of the Bent Pyramid incorporated a spectacular alignment to the sun at the winter solstice, which is seen to disappear behind the pyramid for an observer looking along the longest section of the causeway from the Valley Temple (Belmonte and Shaltout 2009). The Red Pyramid's causeway was instead, probably, orientated due east.

Although many think that the Red Pyramid was constructed because the Bent one was in danger of collapse, a series of hints show that the Snefru project at Dahshur comprised from the very beginning the construction of both pyramids. A detailed analysis of this issue and the critique of the "collapse" theory can be found elsewhere (Magli 2013). Here I only mention that duality in the funerary cult is not a novelty; a striking example is in Djoser's Step Pyramid complex. Djoser's complex has two almost identical underground tombs – one under the pyramid, the other along the south boundary - and duality is apparent in all aspects; in particular, there is a continual, almost obsessive reference to the king as ruler of the "two lands" (upper and lower Egypt). The presence of a unitary Snefru project of the Dahshur landscape – where the southern pyramid has to be interpreted as a symbolic tomb, a cenotaph - is hinted at by the duality apparent in the site – two enormous pyramids of equal height, two (and not three) slopes, two funerary apartments in the south pyramid, two complexes of annexes with the Valley Temple of the "abandoned" Bent Pyramid showing no signs of having been left unfinished or non-functioning. Moreover, a further symmetry was in play in the slopes, since the pyramidion of the Red pyramid has the same slope as the lower part of the Bent pyramid, so that the two pyramids had to some extent specular slopes. What apparently breaks the duality of the project is the presence of a single satellite pyramid. However, this actually helps to prove that the project was a global one, because the Red Pyramid would have had its own satellite if the "machinery" of the king's burial and afterlife was transferred to the northern complex. Finally, the existence of a unitary project is confirmed by the name of the two pyramids. Before Snefru indeed the funerary complexes of the kings had associated "estates" bringing a special name which related the king with Horus. Starting from Snefru it will be the pyramid to have her own name, easily recognizable as such in inscriptions because it is always written in the form <king's cartouche> <text> <pyramid sign>. So the Red Pyramid was called

from the left, the Snefru cartouche, the hieroglyph 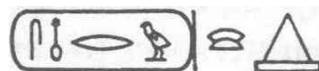-*kha* (meaning shining, rising) and the pyramid determinative. So the Red Pyramid was called *Snefru shines*. The name of the Bent Pyramid was *the same* with the addition of a sign denoting "south", and the whole Snefru complex at Dahshur was referred to as

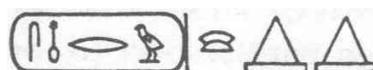

that is, *The two pyramids - Snefru shines.*

The visual effect resulting from such a huge, double project is particularly striking when viewed from the (already old and revered at Snefru times) necropolis at Saqqara, with the two pyramids forming a giant version of the symbolic hieroglyph

*djew* ⌣ . This sign was associated with afterlife and was already extremely ancient, as it appears already in the seals found in the pre-dynastic tombs at Abydos (Dreyer 1998) .
Of course, if Snefru really had a very huge, double project built for him at Dahshur, then it is at least conceivable that his son Khufu devised to construct a double project also for himself.

## 3 The birth name of Khafra

One of the most prominent Mastaba tombs of the eastern Khufu cemetery – which is reserved mostly for the royal family and relatives - lies at the beginning of the first row, in front of the queen's pyramids. It belongs to a son of Khufu called Khaf-khufu (Kelly Simpson 1978).
Since the name of the prince included the royal name of his father, it is found written as

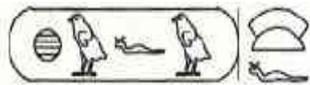

from right to left, Kha-f-[Khufu cartouche]) **(Fig. 4)**. It is, however, doubtful that Prince Khaf-khufu was buried here; besides, his name is not mentioned elsewhere, while many other of the king's sons are known from various sources. Why? The location of the tomb shows that Khaf-khufu was quite an important person during Khufu's reign, and Stadelmann (1984) has proposed a simple solution to this quandary: prince Khaf-khufu was none other than Khafra, who succeeded to the throne unexpectedly on Djedefra's (the first successor of Khufu) sudden death. Upon succession, Stadelmann argues, he substituted the name of the father with that of the Sun God Ra in his name, so that Khaf-khufu became Khaf-Ra:

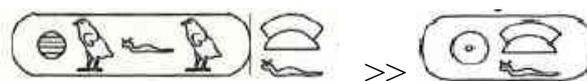

This interpretation has been criticised in that the tradition of assuming a throne name in the royal titulary is well attested only later (Bolshakov 1995). However, independently from any tradition, it is clear that *if* Khaf-khufu ascended to the throne, the Khufu part of his name would have had to be changed to Ra. Indeed, it was impossible for the king to write his own name in a cartouche *including* another cartouche, and on the other hand Khufu probably identified himself precisely with Ra already in his own life, as we will discuss later on.
If Stadelmann's Khaf-khufu = Khafra equation holds true, then of course the newly acceded king hastily abandoned his mastaba tomb; perhaps, then, he appropriated himself of a pyramid already under construction just nearby, in a dock maintained by Djedefra (the cartouche of Djedefra appears in the Khufu boats' pit, showing that Khufu's first successor continued the works at Giza while building his own pyramid at Abu Roash). It is very important to stress that appropriation of royal monuments is

*not* a common wisdom in the Old Kingdom (while it would become extremely common in the New Kingdom), but special reasons – of opportunity, or legitimization of power – may have influenced such a choice. Therefore, we should not necessarily think to the New Kingdom appropriation processes (monuments usurped, with the cartouche names of the Pharaohs overwritten) but rather to a respectful instalment of Khafra's tomb in the south pyramid of the complex of his revered father.

## 4 The satellite pyramids of G1 and G2

The satellite pyramid of G1, discovered in the 90' (Hawass 1996), was added in haste to the existing project and located along the east side of the main pyramid. On the contrary, the satellite pyramid of G2 is in the "canonical" position as defined in the project of Khafra's grandfather Snefru: to the south of the southern pyramid. Why? A solution might be that the Khufu project of course already included a carefully planned satellite pyramid, located *to the south of the southern pyramid*: precisely the one we know as that of Khafra. As soon as the south pyramid – together with her satellite - were commandeered by Khafra, the need for a satellite for Khufu arose, and the ruling king ordered the construction of another satellite for his father near G1.

## 5 The Khafra's Queens

The pyramid of Khufu has three subsidiary pyramids, almost certainly the burials of queens, and the same holds for the pyramid of Menkaura. Curiously, the G2 complex has no such pyramids. From the archaeological records little is known about Khafra's queens, and even less about their burials. A queen called Meresankh III was almost certainly Khafra's wife (although the name of the husband is not attested). She was buried in a mastaba tomb of Khufu's eastern cemetery which was originally prepared for her mother, Hetepheres II. Another of the Khafra's Queen was probably Khamerernebty I, mother of King Menkaura. Her burial was perhaps meant to be the so-called Galarza tomb in the Giza Central field, thus on the plateau in front of the second pyramid. However the tomb was finished for her daughter Khamerernebty II, and the identification of her tomb remains uncertain. A third Queen was perhaps Persenet, again buried in the Central Field, but also in this case the marriage is uncertain. In any case, the question remains, why Khafra did not order to include pyramids for his queens in the pyramid complex? A reasonable possibility is that it was considered unsafe to attempt the addition of subsidiary pyramids on the artificial terrace which is located directly in front of G2. Of course, this is not a proof that the terrace already existed in Khafra' s times; however, how it is possible that an overall, satisfactorily solution - *whatever* it was, for instance a carefully planned row of dedicated, prominent mastabas in the central field - was not planned for Khafra's Queens? The unique answer seems to be, again, that the original project of the annexes of G2 was not meant for Khafra.

## 6 The original project of G2, part 1.

The internal structure of G2 is relatively simple (**Fig. 5**). It consists of a granite-lined descending passage (we will refer to it as passage A) which starts on the north face, becomes horizontal at the level of the rock surface, and leads to the funerary chamber, which was excavated in a pit and then covered with a pent roof. The chamber still contains the black granite sarcophagus, half-sunk into the paving. The descending passage is today closed and used for a ventilation plant, so it is impossible to visit it. I really do hope that the scholars who studied it in the past checked carefully that no joint with a hypothetical plugged ascending corridor similar to that of G1 exists on its ceiling; assuming that such a corridor does not exist, the second pyramid has no internal, above-ground structure. Another underground part exists, anyway. It consists of a second descending passage (passage B), excavated in the bedrock in front of the pyramid. This has prompted the suggestion that the pyramid was originally intended to be more elongated to the north. All the hypotheses of different plans which have been made, however, assume that this second passage was excavated *first* so that the pyramid was originally designed in such a way that the passage crossed the body of the monument (Becker 2004). In one hypothesis, the original project was very small and the second chamber was the burial chamber of a small pyramid whose north face was advanced to the north; in another hypothesis the pyramid was meant to be enormous by any stretch of imagination, with a base side of 470 cubits/246.5 meters, to be compared with the actual 410 cubits/215.3 meters and with G1, 440 cubits/230.3 meters.

There is, however, *no proof whatsoever* that passage B was built before passage A. If this hypothesis is cancelled, then another solution emerges. If we suppose that G2 originally was the south tomb of Khufu, then it was a cenotaph and did not need to have a Serdab, the chamber devoted to house a statue of the king where the rite of the Opening of the Mouth had to be performed. This chamber was indeed already present in the north pyramid (it is the so-called Queen's chamber). Therefore, the original project of G2 included only passage A leading to the main chamber. When Khafra claimed the pyramid for himself, the builders had to add a Serdab, which was clearly impossible to excavate at the corridor level, in contact with the body of the monument. The unique solution was to detach a descending corridor and to go underground to excavate a new room, which for this reason was shaped with a "tent-like" roof imitating that of the Serdab of G1 that is, the Queen's chamber. Clearly, excavating the room starting only from the inside would have been quite difficult, and for this reason – I propose – not one but two new service corridors were excavated, one going down and north from the existing corridor and the other descending from an entrance *outside* the pyramid already in construction.

## 7 The original project of G2, part 2.

There is another strange fact about the interior structure of G2. As in all 4$^{th}$ dynasty pyramids, the entrance corridor is displaced to the east with respect to the mid line However, the corridor of G2 is so displaced, that as a result the projection of the apex does not fall inside the perimeter of the apartments (**Fig. 6**). This is very unusual for an Old Kingdom pyramid. The huge enlargement of the pyramid considered by previous authors to 470 cubits and discussed in the previous section is however so big, that the chamber would not be placed under the apex as well. Analysing an intermediate situation one discovers that, if the original project had a base slightly smaller than the base of G1 (to fix ideas, of 420 or 430 cubits) then the projection of the apex would lie in the chamber. Further, the south-east corner would lie very accurately along the line of azimuth 225° which, coming form the temple of Heliopolis on the other bank of the Nile and connecting the south-east corner of G1 with that of the later Menkaura pyramid, furnished the main topographical axis during the development of the whole necropolis (see last section) and is today missed by about 12 meters.

Thus one might suppose that an original project comprising, as the Snefru project in Dahshur, a south pyramid with a base side slightly smaller than that of the north pyramid but a steeper slope was reduced, for reasons of time and material saving, when the complex was adjusted for Khafra. It should be noticed that it was practically not possible to reduce the size more than this (relatively small) amount because the western courses of the pyramid's corners are in many points "fake" as they are sculpted in to a pre-existing rock noll – certainly one of the first works carried out - so reducing the dimension further would have required quarrying a huge amount of rock anyway.

## *8 The attribution of G2 to Khafra.*

The name of King Khufu appears, as is well known, in the "quarry marks" of the so-called Relieving Chambers of G1. As far as the second pyramid is concerned, although there is scarcely any doubt as to who was the king buried in, namely Khafra, the name of this king – or of any other – does not appear. Of course Khafra is present in the valley temple with his famous statues recovered in it, but the first secure attribution to Khafra of the pyramid *in itself* of which we are aware of is given by inscriptions which are later – although only slightly - in relation to the monument (one example is the nearby mastaba of Qar (Kelly Simpson 1976) where the second pyramid is mentioned as *Khafra is great*). So strictly speaking *there is no proof contemporary to the construction* associating the G2 pyramid with Khafra.

## *9 The attribution of the Sphinx to Khafra*

It is usually assumed that the Sphinx, due to her proximity to the valley complex of Khafra, is an image of this Pharaoh. The Sphinx's face however, at least in the modest view of many people including myself, does not resemble at all the face of the Pharaoh who allegedly had it carved. Also in this case Stadelmann has proposed a simple solution, namely that the Sphinx is Khufu (Stadelmann 1985). *If* this holds true, however, why did Khafra asked for a gigantic statue resembling his father in his *own* complex?

## 10 The "star shafts" of G2

As is well known, G1 has four channels which cross diagonally the body of the monument starting on the north and south faces of the queen and king chamber respectively. The dimension of the channels is very small (around 20 cm) and their function is symbolic: the upper ones exit on the north and south face and connect the "soul" of the Pharaoh with the pole star of the epoch and with Orion respectively, while the lower ones point to Ursa Maior and Sirius but end in the nucleus with two "doors", almost probably symbolic gates resembling those of the Pyramid Texts, where the above mentioned "stellar" destinations are repeatedly mentioned (Badawy 1964, Trimble 1964, Bauval 1993, Magli and Belmonte 2009).
The main room of G2 has two rectangular openings similar to those of the King chamber, which were cut in the upper part of the walls (as mentioned, the room is excavated in the bedrock, only the ceiling is made of stone slabs) **(Fig. 7)**. The corresponding "shafts" were cut for a few tens of centimetres. Their interpretation is very controversial. One possibility is that they were sockets for the insertion of a cross-beam to support a curtain, thereby dividing the room into two compartments, a solution that looks very unlikely. Another possibility, which I actually feel plausible, is that they were sockets used to fix an helping device for works in the room, e.g. moving the sarcophagus in position. However, there is yet another solution, namely that they are star shafts similar to those of Khufu (Edwards 1981). As such, they certainly are "unfinished" or "symbolic" but, equally certain, it would have been impossible to "finish" them - even for the 4th dynasty architects - crossing the whole body of the enormous pyramid. So, it is conceivable that they were added – more or less up to their possible extent - cutting into the bedrock when the sarcophagus room changed its function from the "south tomb" of Khufu to the burial chamber of Khafra.

## 11 The solar alignments

The project of the causeways and of the valley temples of G1 and G2 includes a series of solar alignments **(Fig. 8)**.
Those of G2 are as follows. The causeway slopes down straight from the funerary temple to a point, which we shall indicate as O', located at the north-west corner of the valley temple, reachable from the inside of the building through a spectacular corridor cased in granite; over this point there also passes the ideal prolongation of the southern side of the pyramid. Three alignments related to the cycle of the sun from the spring equinox to the autumn equinox connect this point (and, slightly more generally, the area in front of the Sphinx temple) to the western artificial horizon which was formed by the construction of the two great pyramids. First of all, a line running due west passes along the south side of the pyramid, and so the sun at the equinoxes was (and is) seen setting in alignment with the south-east corner on those days. Second, the alignment defined by the causeway is oriented ~14° north of west. The azimuth of the setting sun at the summer solstice at the latitude of Giza is ~28° north of west, and therefore this alignment coincides with the midpoint of the path of the setting sun at the horizon between the spring equinox and midsummer and between midsummer and the autumn equinox (Bauval 2007). Finally, let us consider the line which points towards the midpoint of the segment, separating the south-west corner of the two pyramids. The azimuth of this line is twice that of the causeway, ~28° north of west, and therefore it coincides with that of the sun at the summer solstice. Consequently, the midsummer sun is seen setting in between the two pyramids **(Fig. 9)**.

The symbolism implicit in this alignment was noticed for the first time by Egyptologist Mark Lehner during his fieldwork at the plateau (Lehner 1985b). He realised that, when the midsummer sun sets, an observer actually witnesses the formation of a spectacular replica of the hieroglyph 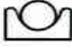 -*Akhet*. This sign represents the sun setting (or rising) between the two symbolic, paired mountains of the sign 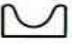 .

This dramatic phenomenon is thus a hierophany, a manifestation of divinity which occurs in dependence of a celestial cycle. Indeed the word Akhet, usually translated simply as "horizon", had a profound symbolic meaning for the ancient Egyptians. The region called in this way (written phonetically *Akh-t*, see discussion in last section) in the Pyramid Texts has the same root as -*Akh*, spirit, and denotes the place where the dead were transformed, preparing themselves for the afterworld. As part of the sky, it was also the place in which the sun, and hence the king, was destined to be reborn. Interestingly enough, as we have mentioned, since pre-dynastic times the symbol of the double mountains was associated with the afterlife, and perhaps this was the message of the double Snefru project at Dahshur as seen from the already revered necropolis at Saqqara. By connecting the double mountain horizon with the sun, an even stronger image associated with rebirth was created at Giza. Clearly, the choice of the summer solstice for the date was not coincidental, since it heralded the flooding of the Nile, and thus the theoretical beginning of the flood season.

Let us now analyse the solar alignments present in the G1 complex. This analysis will be (of necessity) more speculative given that the valley temple is unexcavated and lies under the houses of the modern village of Naziet. In the course of the years however, probes operating there (Messiha 1983; Goyon 1985; see also Lehner 1999) demonstrated the existence of an important building, if not the valley temple itself, located in the area where the ideal prolongation of the northern side of the pyramid and the causeway intersect each other, at a point which we shall denote by O. Clearly, such a point – analogue of point O' – played a special role in the plan of the complex. By drawing from this point a set of lines specular to those of O', we see that they are related to the cycle of the sun from the autumn equinox to the spring equinox. Indeed, a line directed due west passes near the north side of the pyramid, and therefore the sun at the equinoxes was seen setting in alignment with the northern corner on those days. The alignment defined by the causeway is oriented 14° south of west, and therefore coincides with the midpoint of the path of the setting sun at the horizon between the autumn equinox and midwinter and between midwinter and the spring equinox. Finally, the line of the setting sun at the winter solstice (~28° south of west) is directed towards the second Giza complex and passes near the centre of the funerary temple. As a result, the midwinter sun is seen setting beyond the second pyramid (Magli 2008).

Taken as a whole, the Giza complexes G1 and G2 have *symmetric* solar alignments. In particular, the two causeways have opposite azimuths with respect to the east-west line – which shows that the direction of the G2 causeway was chosen accordingly to a symmetry criterion – and each complex embodies a hierophany at a different solstice, a hierophany which, however, can only manifest itself in the presence also of the pyramid of the other complex.

Also in this case, the hints to a unitary project of the two complexes are self-evident.

## *12 The orientations*

All Egyptian pyramids of the fourth dynasty are orientated in such a way that the sides of their bases run parallel to the cardinal directions with very high precision. The errors are so low that to measure these deviations today, we need to use very accurate instruments such as a transit (theodolite) surveyor or high-precision global positioning system (GPS).

Exact measurements of pyramid orientation were first taken by Petrie (1883) and more recently by Dorner (1981) and Nell and Ruggles (2013); rounding the data to the nearest arc minute and considering them within the resolving power of the human eye, ±2', which clearly was the physical limit to which the original builders were anyway constrained, we can fix our ideas using the following average values (Belmonte 2001; for data on each side separately see Nell and Ruggles 2013):

1. Pyramid of Meidum: –18';
2. Bent Pyramid: –12';
3. Red Pyramid: –9';
4. Pyramid of Khufu: –3';
5. Pyramid of Khafra: –6';
6. Pyramid of Menkaura: +14'.

I have omitted from the list Djedefra at Abu Roash – which is in poor condition and not, in my opinion, measurable with a comparable degree of accuracy (in spite of contrasting claims; see Mathieu 2001) – and the "Great Pit" at Zawiet el Arian, which regrettably has never been measured.

Clearly, these are orientations which were obtained with a maniacal accuracy. Although it is written in many sources that it is possible to achieve it with methods based on the sun (see e.g. Isler 2001), no one has ever been able to obtain similar results using solar methods in modern times, and I strongly believe that it is, indeed, impossible to measure shadows with the required accuracy. Thus, by far the most likely candidate for the method used by the Egyptian surveyors is a stellar method, and in particular, the so-called simultaneous transit (Spence 2000 – see also Rawlins and Pickering 2001; Belmonte 2001b; Magli 2003, 2009a).

It consists of observing two circumpolar stars. The surveyors kept track of their relative positions and identified north as the direction on the ground corresponding to the segment joining the two stars when it is perpendicular to the horizon. If the precessional drift of the segment is calculated as a function of time, the resulting graph is a straight line. On this line, the experimental points corresponding to the orientation errors of the pyramids can be positioned, and dates of construction can be read on the time axis. If the stars were sighted with one in the upper culmination and the other in the lower one, then they were most probably Kochab – a star we have already met – and Mizar (see Fig. **10**). The corresponding chronology emerges, however, as being a little earlier than that usually accepted. Higher chronologies (to fix ideas, with Khufu's accession in 2550 BC) fit better with the use of two stars sighted at the same (upper or lower) culmination: Megrez and Phecda.

The simultaneous transit method is the only one ever proposed that takes into account the effect of precession on the distribution of the orientation errors and, at the same time, is compatible with what we know about the ancient Egyptians' way of thinking and about the instrument they used, the so-called Merkhet (essentially a plumb line with a viewfinder).

But there is a problem.

Looking at the true data (instead of looking at their absolute value), it is clear that the measurements made for the orientation of the second pyramid were made before, and not after, those made for the first (Magli 2003). A workaround for this problem is to find some way of "changing the sign" of the value of G2: if indeed it becomes +6' then the "distance" in arcminutes between G1 and G2 is of 6'-(-3')=9' and this gives the sought distance in time between the two orientations, say some 25-30 years.

These ways actually exist: in Spence's approach one has to suppose that, for some unknown reason, the orientation ceremony of the second pyramid took place during summer, while that of all the others was carried out in winter. Similarly, in Belmonte's approach one has to admit that, for some unknown reason, for the Khafra pyramid the orientation was made observing the two stars at their lower culmination (i.e., "under" the pole), while for all the other pyramids the upper culmination was used.

At least in my view, both solutions are definitively *ad hoc* and, as such, definitively unacceptable from the physical point of view. Further, they are not in agreement with what we know about the way of thinking of the ancient Egyptians. The foundation of a sacred building was indeed a complex ritual, in which the king performed a series of symbolic acts such as, for instance, the laying of the foundation's deposits and a ceremony - called the *Stretching of the Cord* - which was so important as to be represented in an equal manner during 3 millennia of Egyptian history. So the orientation of the king's pyramid must have been an operation – a ceremony indeed – of the utmost significance. Therefore, it is likely that it would always have taken place under the same conditions. Furthermore, orientation data would be scattered more randomly if the season of foundation, or the choice of the star's configuration at culmination, was random, so again, it is necessary to admit facts which look illogical for, and only for, the Khafra complex.

There must be another solution, and the natural solution comes from the fact that we are speaking of naked-eye observations, although aided with simple instruments, aimed at orienting huge monuments. In this context, as first noticed by Juan

Belmonte, insisting that the data for the first and the second Giza pyramids seen as they stand (without *ad hoc* change of sign) are separated by only 3' is useless. Much more reasonable is the idea that the two orientations were performed at the same time (Shaltout, Belmonte and Fekri 2007b; Magli 2005, 2009; Magli and Belmonte 2009). Logically, it follows that two projects were laid down simultaneously.

***

To sum up, there exists as much as 12 clues casting doubts on an original Khafra project. Some of these are at least intriguing: the lack of stellar shafts, the double entrance and the lack both of subsidiary pyramids and a well-organized Khafra's queens cemetery, the lack of resemblance of Khafra with the Sphinx. Others actually do cry out for an explanation: the fact that Khafra could build his new tomb in the most favourable position of the plateau because Khufu for some unknown reason discarded the same position, and that he could cut out his own causeway from the bedrock without constructing it because Khufu's stonemasons for some unknown reason left it as a boundary of their own quarry; the fact that Khafra built his own satellite pyramid in the same position of that of his grandfather Snefru while his father's architects forgot endowing the great Pyramid of a satellite and hastily built one in a odd position later on, and finally, the data showing that the Khafra astronomers made their measures to orientate the pyramid during the opposite season (or using two stars at the opposite culmination) with respect to the custom of all previously done orientations at Dahshur and Giza.
Actually, the amount of evidence seems overwhelming for a project devised by Khufu, carried on by Djedefra and re-assessed for Khaf-Khufu, who unexpectedly accessed to the throne.
But it is not enough yet, since we have the very last piece of the puzzle to consider.

## 3. The Giza written landscape

Writing is documented in Egypt since about 3300 BC. The introduction of hieroglyph writing was related to the increasing complexity of regional interactions and the formation of the state (see Regulski 2010 and references therein). From the very beginning, it developed as a mixed system. A component was phonetic, including alphabetic glyphs, another was logographic, representing morphemes; the texts also contained determinatives, images which were not usually read but helped in specifying the meaning of words; glyphs could be used with ideographic meaning on occasion.
The meaning of the word (Greek-derived) "hieroglyph=sacred sign" is actually well suited to express the profound connection of writing with religion. This connection is related to cosmological beliefs, for the very act of creation in the Egyptian conception appears to be connected with writing in a structural analogy between language and cosmos (Assmann 2003,2007). This analogy is based on a one-to-one relationship:

the totality of creation is made of: "all things, all hieroglyphs." A written sign was thus connected with a real thing in a manner somehow similar to the relationship between thing and concept later elaborated in Greek philosophy.

Many hieroglyphs were connected with sacred architecture, and it is therefore natural to investigate on the role of the above mentioned analogy in the architectural context. Well, it is actually a strange experience to realize that Egyptian sacred architecture is plenty of gigantic *replicas* of hieroglyphs in a complex and – as I will endeavour to explain - *interactive* way.

First of all, an architectural entity can be a physical realization of an idealized, but easily recognizable object. In particular, a pyramid is a giant actualized replica of her own hieroglyph 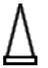, and the same holds for an obelisk 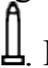. However, let us consider the hieroglyph 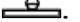. It represents a stylized loaf of bread on a reed mat, meant as an offering scenario. Thus, we could manage to guess that the hieroglyph means "altar", and indeed, it is so. Well, Egypt is plenty of *stone* altars which are giant replicas of the hieroglyph which *represents* an altar. The most spectacular I have ever seen is certainly the quadruple alabaster altar orientated to the cardinal points which occupies the court of the 5th dynasty Sun Temple of Niuserra **(Fig. 11)**, but hundreds of examples could be cited. Which one came first, the stone altar or the altar glyph? This is probably an example in which architecture replicates in a *purely symbolic* way a hieroglyph which in turn was invented trough a symbolic, cognitive process. An interaction thus exists between the generation of writing icons and that of tangible, architectural things. Another example are the fifth dynasty tombs of officers in Saqqara south which appear to have a *plan* resembling the hieroglyph group 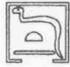 - *per-djet* (house of eternity) (Dobrev 2006).

These examples show that, in ancient Egypt, invention in sacred architecture and invention in sacred writing were mutually interchangeable. Interestingly enough, this interplay can be brought to yet an *higher* level of sophistication, that of the *written landscape*.

To explain what I mean, I first need to recall the concept of sacred space. This fundamental concept arises from the fact that, although space belongs to man, man's space is of course a place indifferent to human thoughts and feelings, something that strikes us as human beings as being disagreeable. So, the possibilities of breaking away from uniformity – whether they are natural features of the landscape, such as rocks, rivers, caves, mountains and so on, or man-made features – help in identifying the place where a group of humans lives as *the correct* place to live trough a process of religious foundation. The same happens when looking at the "third dimension"- the sky. As a consequence, uncountably many civilizations have looked at the celestial realms and cycles, framing them together with human space in the human worldview, the *cosmos*.

Cosmos gives order, and order transforms space into what we call *sacred space*. Sacred space is therefore the final result of the "cosmisation" process: a place which has a centre and which has been founded, ordered and prepared for human life. The historian of religion Mircea Eliade (1959) was probably the first to understand this mechanism in detail when he wrote:"The Experience of Sacred Space makes possible the founding of the world: where the sacred manifests itself in space, the real unveils itself, the world comes into existence."

A tremendous power is of course tied up with the cosmos, and so each sacred space, or sacred landscape, is also a *landscape of power*. In a sense, the construction of a "cosmovision" is also the foundation of temporal order, that is, again, power. Further, power is associated with religious ideas on afterlife, which again were connected with sacred spaces and celestial cycles, in particular, in many cases, with the cycle of the sun and therefore with the solstices. In most cases, it was the winter solstice, when the rising sun at the horizon stops moving to the south and the length of daylight ends its decreasing, to be singled out. This is not, though, the case with Egypt. Indeed in the arid climate of Egypt the main natural phenomenon that allowed the "rebirth" of the crops was the inundation of the Nile. This is a gradual phenomenon, but usually begins at the end of June. Therefore, it is rather the *summer* solstice which was identified as the harbinger of the renewal/rebirth; at that time, also the concomitance with the Heliacal rising of Sirius was noted and the beginning of the Egyptian calendar fixed.

Once the cosmic order was established and ways to measure it devised, the need to control its correct, cyclic evolution arose. Here, monumental architecture definitively came into play. It is indeed with the man-made, modelled features of the sacred landscape – temples, or tombs, or even entire towns – that the sacred is symbolised and, in a sense, controlled. A sacred place is thus also a point of contact, a place where a *hierophany* - an explicit manifestation of the divine - occurs. In the Old Kingdom, the sacred space was to be a sort of "exclusive" to the king, and since sacred space was bound up with the divine nature of the Pharaoh, the sacred landscapes *par excellence* were also *funerary landscapes*, connected with the Pharaoh's afterlife. Giza clearly is the most astounding example, and indeed in Giza we find the key elements which characterize a sacred landscape: order, orientation, and connection with the celestial cycles. But since in Egypt sacred architecture and sacred writings were, as we have seen, connected, so were sacred space and sacred writings: as a consequence, I propose to define Giza as a *written* sacred landscape. In such a framework, not only the pyramids are giant replicas of hieroglyphs, but all the elements *combine accordingly to "writing" rules* to transmit – actually to write - explicit messages.

A first example is the symbolic meaning of the topographical axis which, running south-west/north-east (azimuth 45°) strictly governs the subsequent placement of the monuments of the Giza plateau. The line runs indeed along the diagonal of Menkaura's first queen's pyramid, touches the south-east corner of Menkaura's pyramid, follows the diagonal of his funerary temple, passes the south-east corner of the second pyramid court, cuts the diagonal of the fore-temple, touches the south-east corner of Khufu's Pyramid and very nearly cuts the diagonal of his first queen's pyramid (Lehner 1985b, Jeffreys 1998). If the axis is prolonged across the river, it points in the direction of the temple of Heliopolis. Heliopolis housed the most important temple of the Sun God Ra. It was also a primary theological centre, where the dominant cosmological doctrine of the Old Kingdom, the Great Ennead, was formulated. In a nutshell, this doctrine identified the ascendence of Osiris from the creator version of the Sun God, and since the Pharaoh was identified with Osiris' son Horus, it underpinned the divine nature of the monarch. Little remains today of the Heliopolis sacred centre, whose ruins lie under the modern buildings of Cairo in a district known as Mataria. However, one obelisk (erected by Senwostret I, the second Pharaoh of the twelfth dynasty) remains standing in (or near) its original position. It is therefore a useful survey point and, as a matter of fact, when the Giza axis is projected using a satellite mapping software like Google Earth, it passes very close to the obelisk (the distance is considerable but, taking into account earth's curvature, a sun-reflecting signal located in Heliopolis at – say – 20 metres above ground would have been easily visible from the west bank). Thus the Giza axis deliberately pointed to Heliopolis, and the two places, though quite a distance apart, "spoke" each other. About what?

The answer is in the key role of the sun god during the Old Kingdom (Quirke 2001). This god was worshipped in various forms, each associated with a different aspect, such as Ra-Atum, the creator, or Ra-Horakhti, identified with the morning sun. The sun disk itself was considered to be the visible body of Ra, sailing through the sky on his barque, mentioned in the Pyramid Texts as one of the places where the deceased Pharaoh had a designated seat. During the night, Ra entered the underworld, only to be reborn the following dawn. Ra had also, of course, a direct influence on the earth, governing the seasons; this influence was recognized especially in the association of the summer solstice with the flooding of the Nile.

Given such a complex scenario, it is not surprising that the relationship of the sun god with kingship came to be the most significant element of the ideology of power during the Old Kingdom. The point of rupture, the moment when the Pharaohs adopted their solar identification as the true basis of kingship, appears to have been during Khufu's reign (Stadelmann 1985). Several hints indeed indicate that Khufu in some way declared himself to be the Sun God. His sons and grandsons would consequently start to define themselves "Son of Ra" and to acquire explicitly the Ra-particle in their royal names. Since Heliopolis was – at least in a sense – the birthplace of Ra, an explicit, topographical link of the kings' tombs with the sacred centre becomes understandable. This topographical relationship can be defined, therefore, as a *dynastic axis,* and it is perhaps alluded at in a passage of the Pyramid Text (PT 307) where the kings states "My father is an Onite, and I myself am an Onite, born in On when Ra was ruler" (On stands for Heliopolis in Faulkner's translation used here).

As a consequence of the axis, anyone approaching Heliopolis and looking towards Giza in ancient times could only see the Great Pyramid, because the second pyramid (to say nothing of the third) was almost hidden from sight by the gigantic mass of the first. The Giza pyramids thus were meant to *queue after the city of the sun* along the Giza axis. In Hieroglyph writing, it often occurred that God's names appeared as parts of personal names. In these cases they were usually written at the beginning irrespectively from their actual position in the word. Similarly, along the Giza diagonal, we can subsequently "read", from Heliopolis all the way up to the desert to the south-west of the Menkaura pyramid, the "name" of a God – Ra – represented by his temple-house and the name of his direct descendants, subsequent incarnation of Horus on earth. The names of these kings are represented by their respective pyramids, and indeed each pyramid had a specific name associating the monument with his owner; in a sense, the pyramid's name is a sort of further entry to be added to the royal titulary, the list of traditional names which characterized each Pharaoh

A second example of a written element of the sacred landscape at Giza is clearly the Akhet hierophany which is also, as we shall see in a moment, a pyramid name. With the Akhet sign the structural analogy which connects writing with the cosmos and thus with the sacred landscape is brought to yet a higher level of sophistication – notwithstanding the crystal-clear, immediately recognizable and non-esoteric nature of her message. This level is precisely that of the cosmos, since it not sufficient to have two giant pyramids each weighing some 8 millions of tons to obtain the Akhet sign – yes, to *write* it – but also the sun in the correct position once a year is needed. Now, the question is, who was the king responsible for such an astonishing demonstration of ability in mastering the cosmos at his will?

It is very likely to think that he was the same king who *called* his own complex in the very same way. We can learn the name of the complex of Khufu from many sources; for instance, from the list present in Qar's mastaba. It was

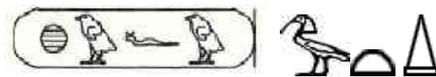

that is, "The Akhet of Khufu", where Akhet is written phonetically as 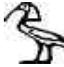-*Akh-t*. Therefore, the name bequeathed to us for the Great Pyramid is *the same* as that of the spectacular hierophany occurring at Giza at the summer solstice. This hierophany replicates – yes, *writes* - the very same name as the Khufu project in the sky once a year. The Khufu complex was *called* Akhet because it actually *was* the Akhet – the symbolic horizon and, trough structural analogy, the hieroglyph with this meaning – pertaining to the divine Khufu, who had been united with the sun god.

Clearly, the complex was designed to include from the very beginning both the "mountains" needed to realize this.

## 4. Discussion

The meaning of the gigantic sacred landscape of King Khufu can be better understood by a final comparison to that of his father. The similarities between the sacred landscape devised by Snefru at Dahshur and the project of the two main pyramids at Giza are indeed striking: two giant pyramids of almost equal height, with two different slopes (as mentioned, the upper section of the Bent Pyramid has the same slope as the Red Pyramid, so that only two differing slopes are present at Dahshur), two sets of annexed buildings, but (originally) only one satellite pyramid to the south of the southernmost main pyramid. There is, however, a key difference. The Snefru project was planned to dominate the southern horizon of Saqqara as a double mountain sign 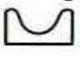, while Khufu chose to transform this symbol into 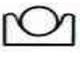, the

Akhet sign, an even more powerful icon of rebirth, by "adding" the sun in between his own paired mountains.

This interpretation has been criticized (Tedder 2009) on the basis that the name of the Khufu complex is, as already mentioned, written phonetically (in the Qar mastaba and in the other known instances). We actually do not know the way in which the name of the Khufu complex was written during Khufu's reign (the unique possible attestation comes from a re-used fragmentary relief where the pyramid's determinative is not present); the "double mountains plus sun" sign first appears during the 5th dynasty. There is, however, no doubt whatsoever about the identity of meaning of the Akhet sign with the Akhet word; the sign will later become synonymous with the king's tomb in general, and with Giza in particular. For instance the cult of the Sphinx established in the 18th dynasty by Amenhotep II will constantly refer to the statue as Hor-em-akhet=Horus in the Akhet, where the Akhet in question, always and rigorously written with the Akhet sign, obviously is the Khufu's one, taking place each year at the summer solstice just to the rear of the huge statue (Fig. 12).

To conclude, thus, it seems likely that the hieroglyph sign entered in use after its "primeval" actualization devised by Khufu's architects at Giza: not only Giza was a written landscape, but it triggered the creation of a new writing sign.

The complexity of the interplay between creation, writing, and architecture can be better understood in the framework of cultural memory: the preservation of collective knowledge from one generation to the next, and the construction of a collective identity trough a shared past (Assmann 2011). In Egypt, cultural identity and religious feeling were deeply connected with the idea of preserving Maat, the Cosmic Order, on Earth. Such an order was anchored to the celestial cycles: the cycle of the sun, the calendar, the succession of the hours of the night, the reappearance of Sirius and so on. It was the regularity and predictability of the Cosmic Order, assured on earth by the existence of a living god, the Pharaoh, that held together the country. For this reason, sacred landscapes such as Giza were ordered and anchored to the celestial cycles as well, and the pyramids functioned both as landmarks and as time marks of collective memory. The messages transmitted by the monuments and by their relative positions in the landscape can be hard for us to understand, but this does not mean that such messages were hidden. On the contrary, they were perfectly tuned and meaningful to the people they were addressed to. In Assmann's words "Cultural memory is a kind of institution. It is exteriorized, objectified, and stored away in symbolic forms".

When such symbolic forms are connected with the repetition of the celestial cycles, their creators can be relatively sure that their messages have been stored for the millennia to come, and indeed, we can still see the rejoicing of king Khufu with the Sun God Ra every year, at Giza, at the summer solstice.

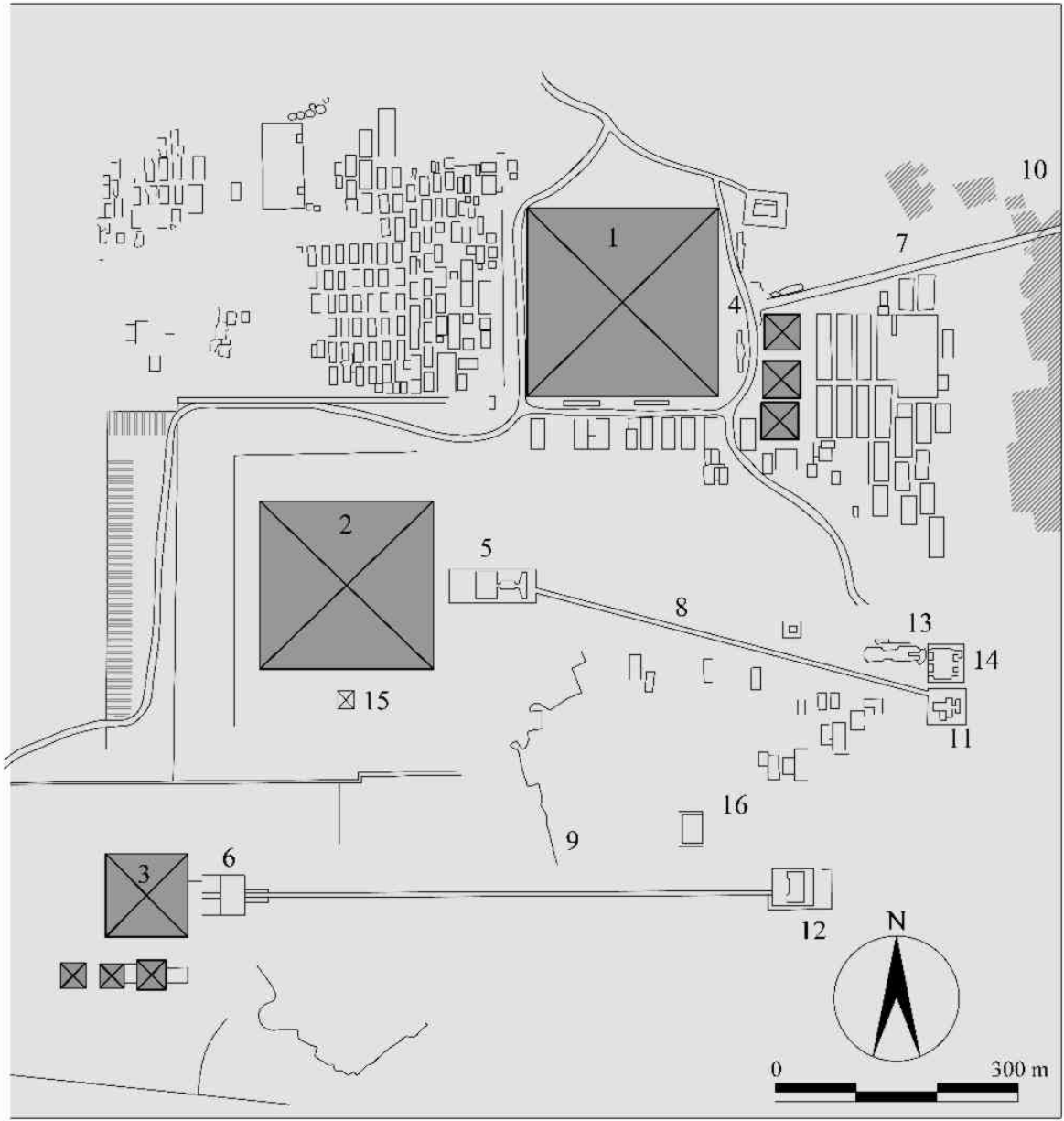

**Figure 1**
Schematic map of the Giza necropolis: (1) pyramid of Khufu; (2) pyramid of Khafre; (3) pyramid of Menkaura; (4–6) funerary temples; (7–9) causeways; (10) modern village; (11) valley temple of Khafre; (12) valley temple of Menkaura; (13) Sphinx; (14) Sphinx temple; (15) satellite pyramid of Khafre; (16) Khentkaues tomb.

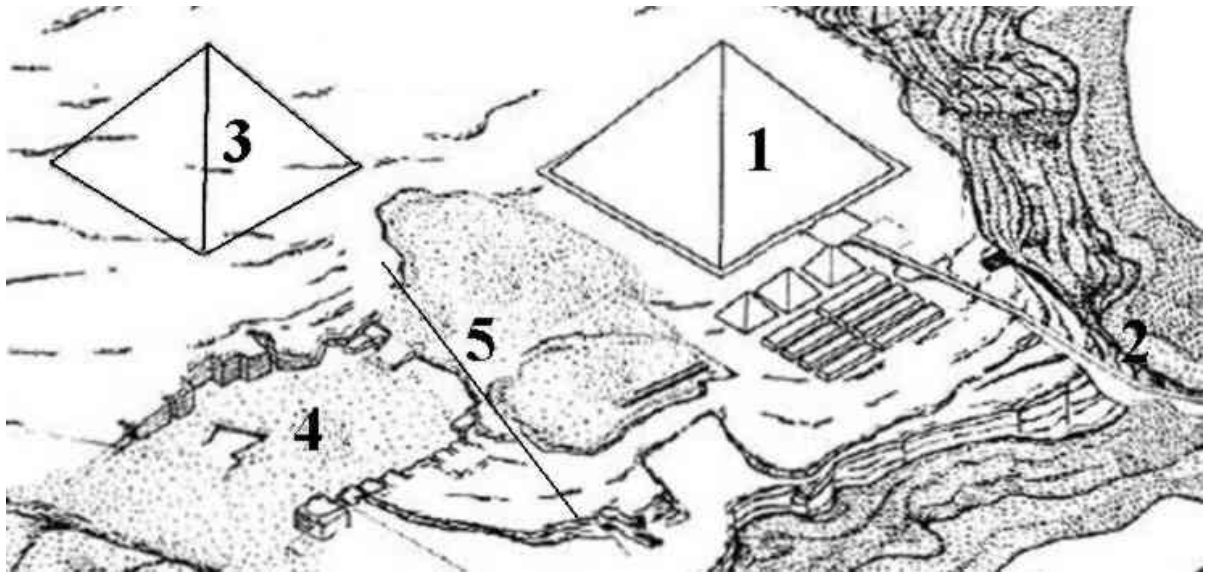

**Figure 2**
A Reconstruction of the Giza Plateau towards the end of the Khufu project. Khufu has selected for his pyramid (1) the rock terrace to the right (north) and in doing this has been obliged to construct a huge causeway leading to the valley (2). In doing this he neglected the favourable area to the immediate south-west where the second pyramid (3) will later be built; in spite of this, his quarry (4) does not extend to the whole escarpment but "spares" a rock ascent which will be later pretty useful for the astronomically oriented Khafra causeway (5). (Drawings by the author, adapted from Lehner 1985).

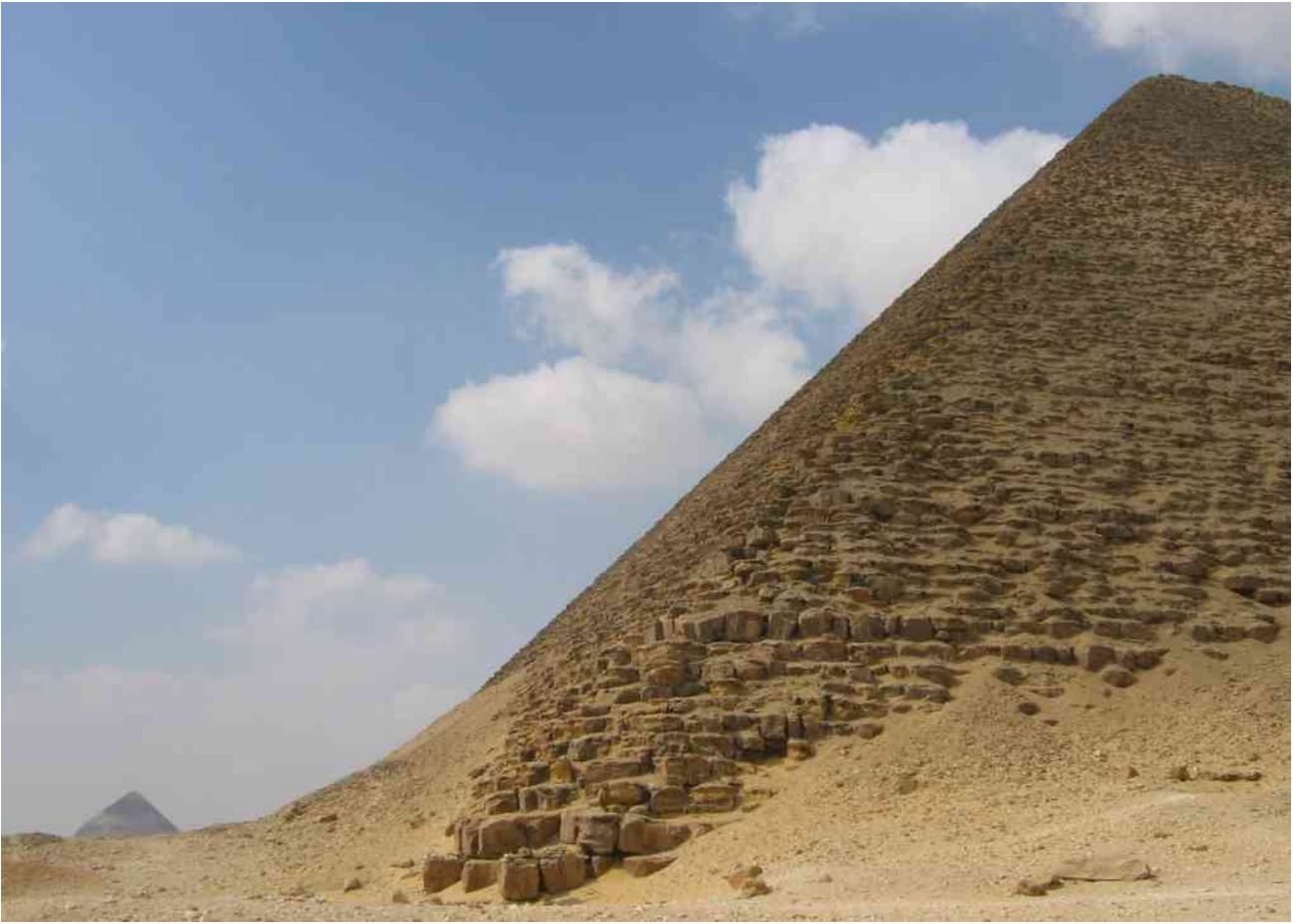

**Figure 3**
**The Snefru project at Dahshur: the Red Pyramid (foreground) and the Bent Pyramid (background). (Photograph by the author)**

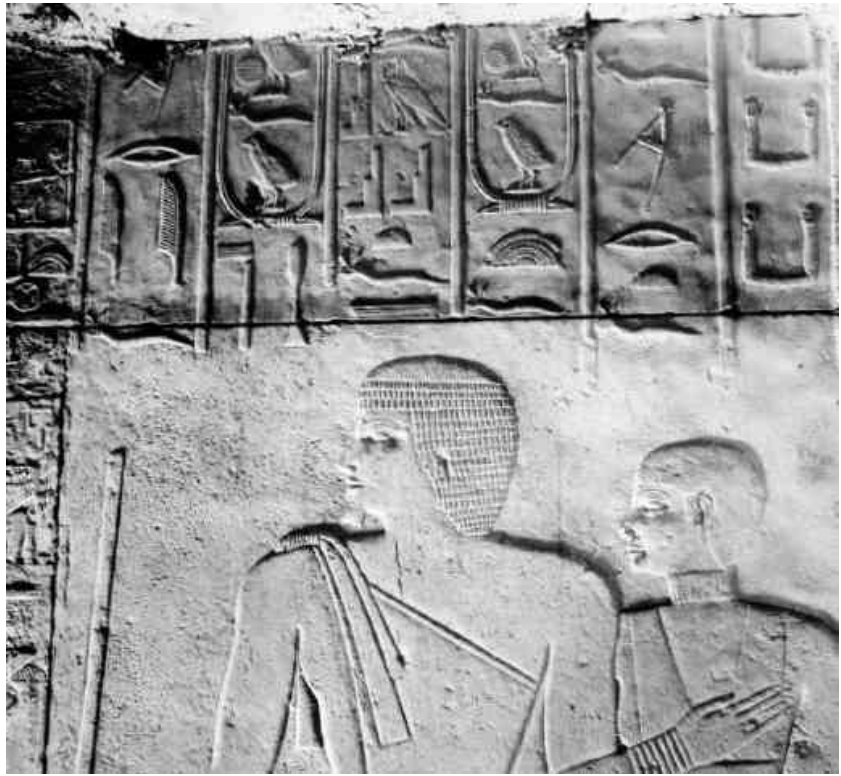

**Figure 4**
Reliefs in the inner chamber of Khaf-Khufu mastaba. The name of the prince is readable in the third register from the right.

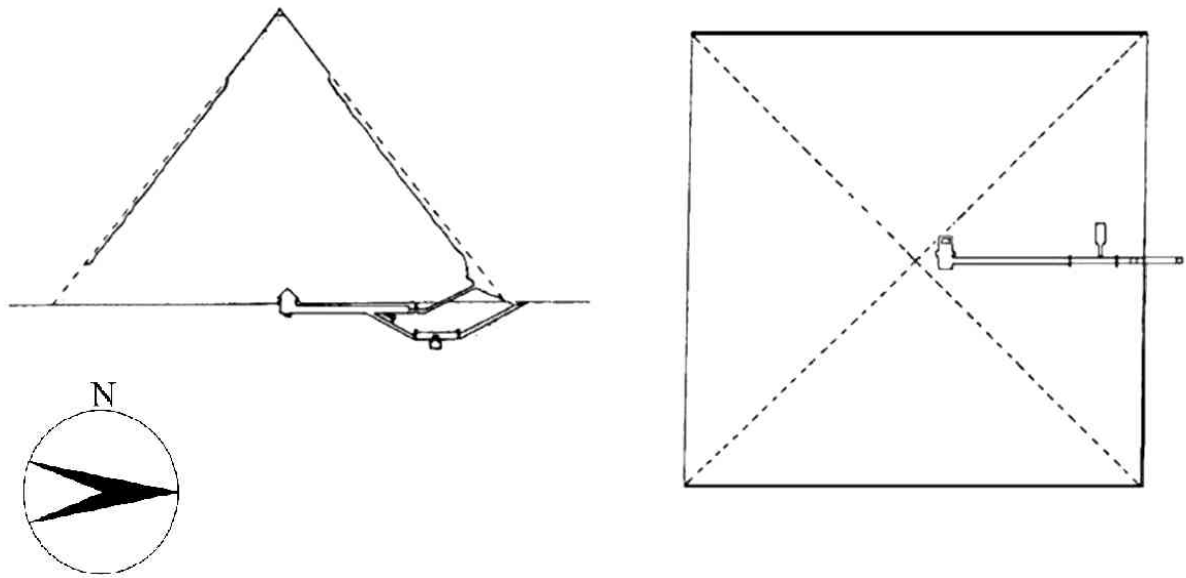

**Figure 5**
**Section and projection plan showing the internal structure of G2**

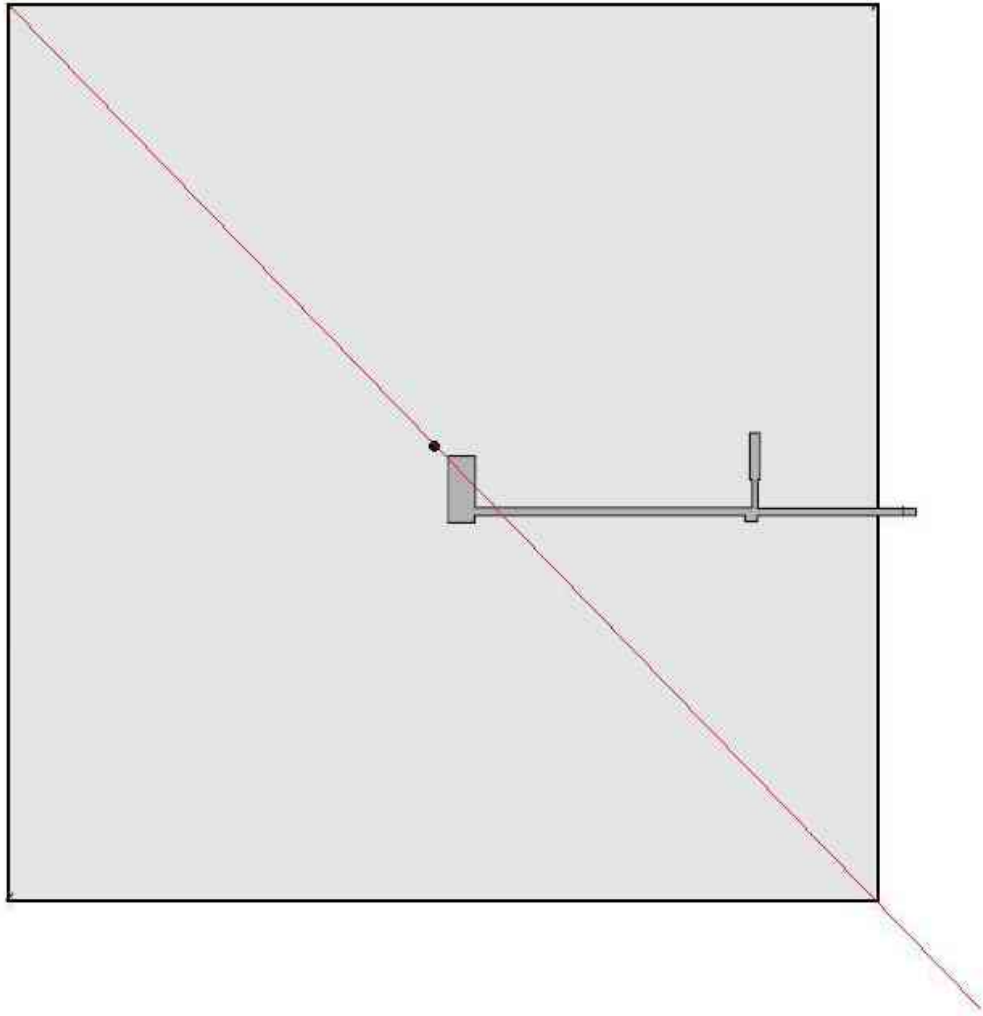

**Figure 6**
Projection plan of the internal structure of G2 (north to the right) showing the the projection of the apex as a black dot. Today it is definitively out of center. If the original project of the pyramid was larger, the south-west corner (upper right) was in any case fixed and unmovable. Therefore the hypothetical enlarged project can be visualized "sliding" the black dot along the red (south-west/north-east) diagonal. If the enlargement was relatively small (10 to 20 cubits base) then the black dot falls in the burial chamber, otherwise it slides too much and again falls outside.

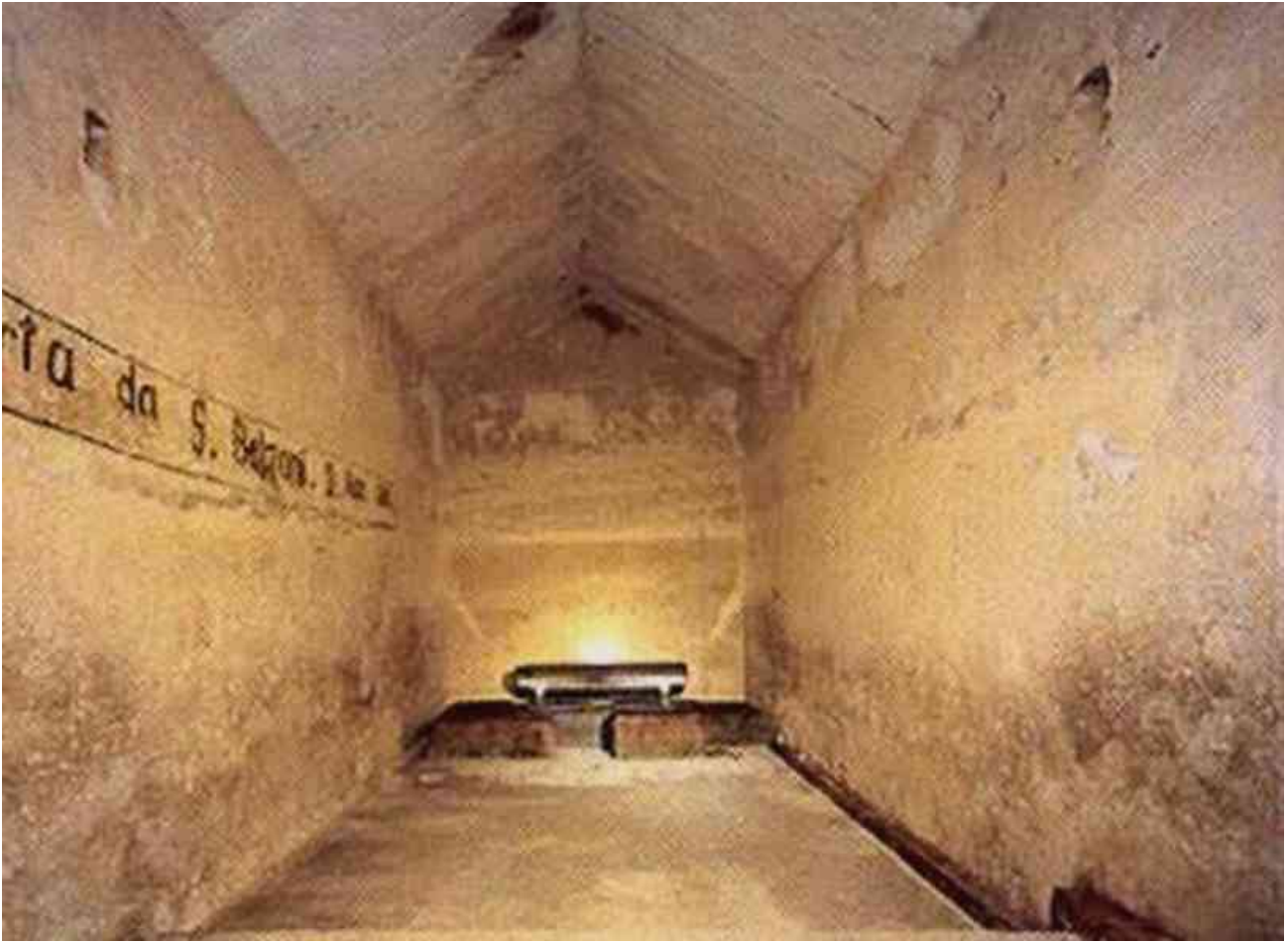

**Figure 7**
**Giza. The main room of G2. (Photograph by the author)**

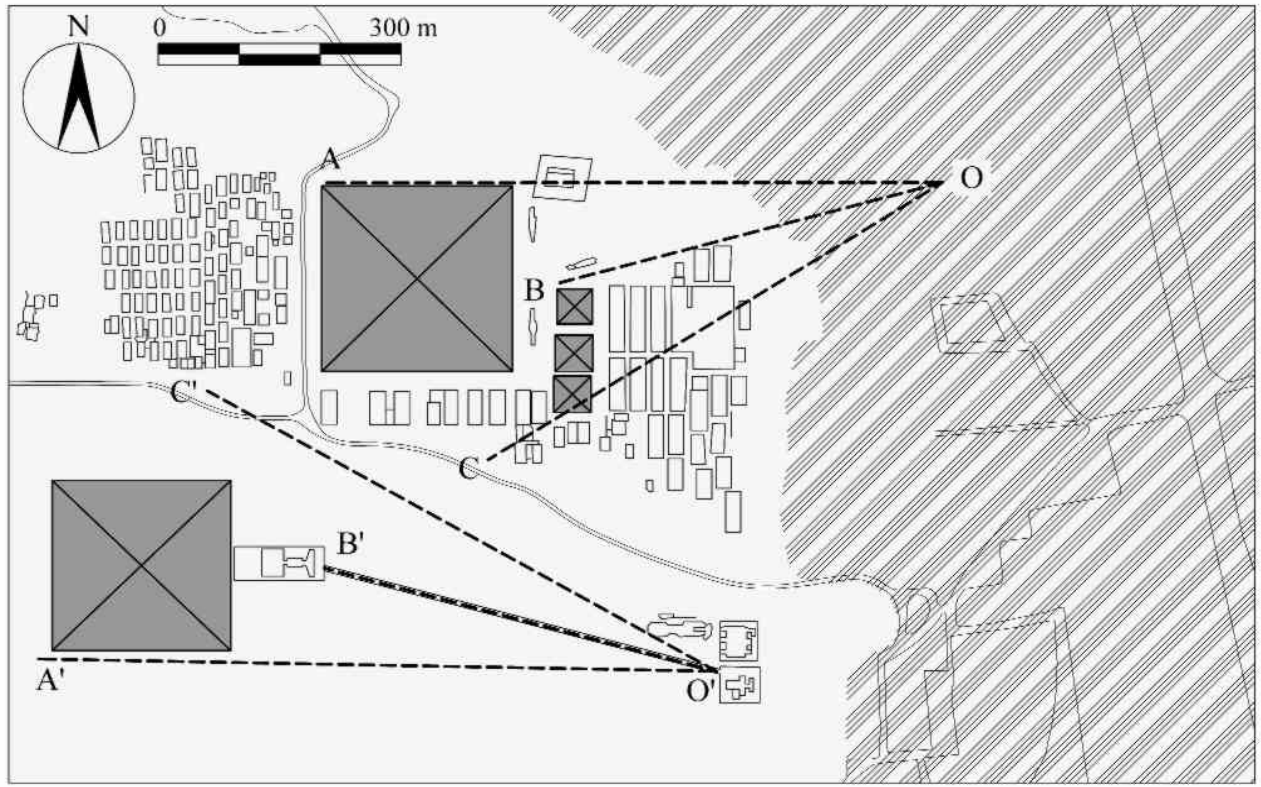

**Figure 8**
**The solar alignments of the two main pyramid complexes of Giza.**

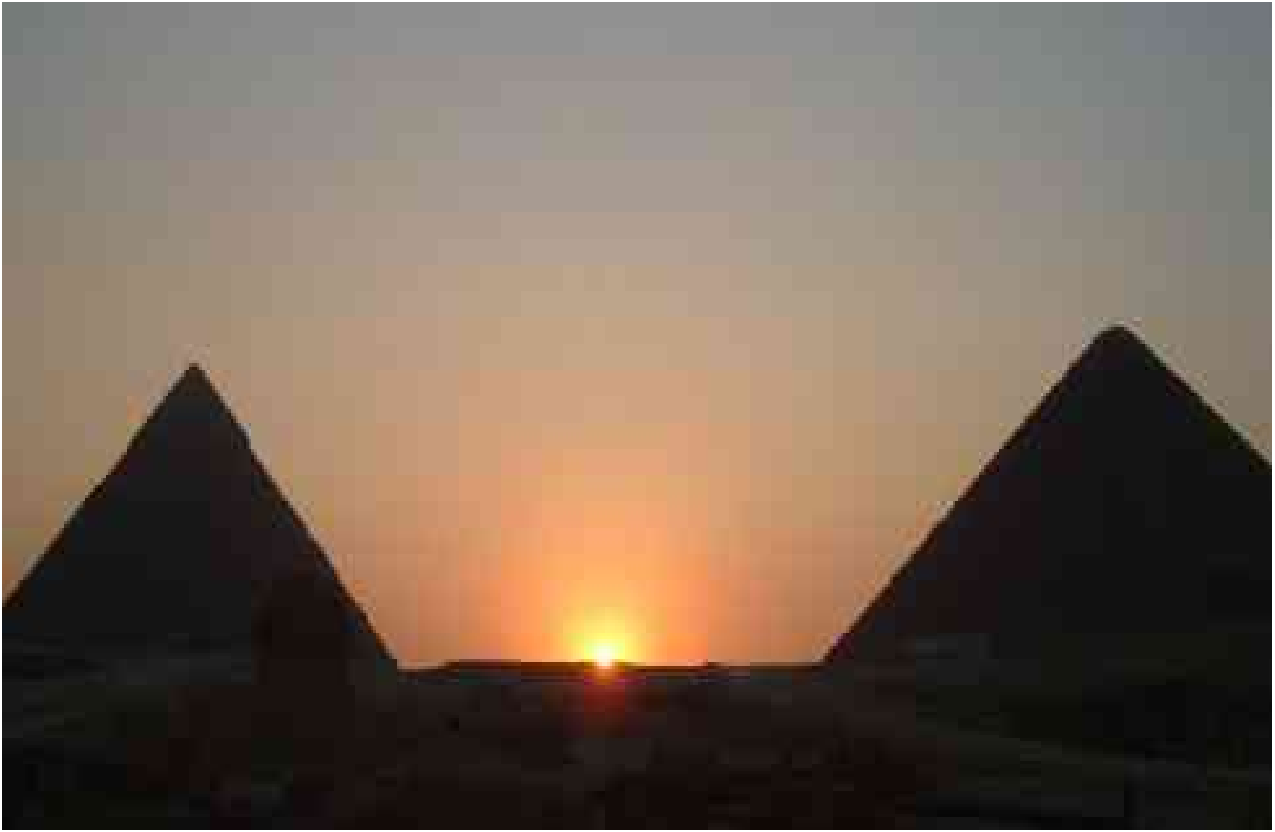

**Figure 9**
**Giza, summer solstice 2011. Akhet, the name of the Great Pyramid, is rewritten once a year from 4500 years on the Giza plateau and is composed by two enormous stone monuments and the sun setting in between them. This picture was taken from the terrace in front of the Sphinx.(Photograph by the author)**

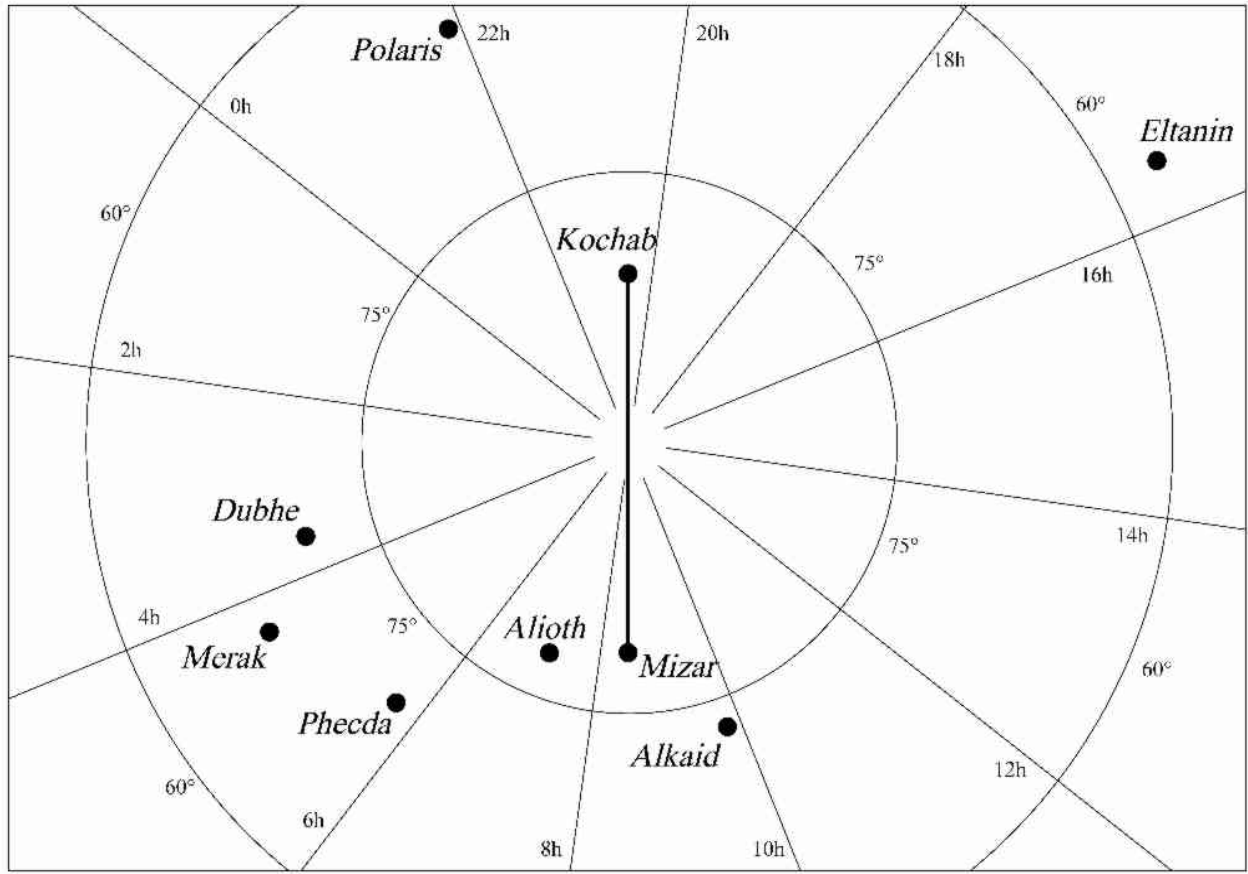

**Figure 10**
**The simultaneous transit of Kochab and Mizar in the northern sky of Giza around 2500 BC.**

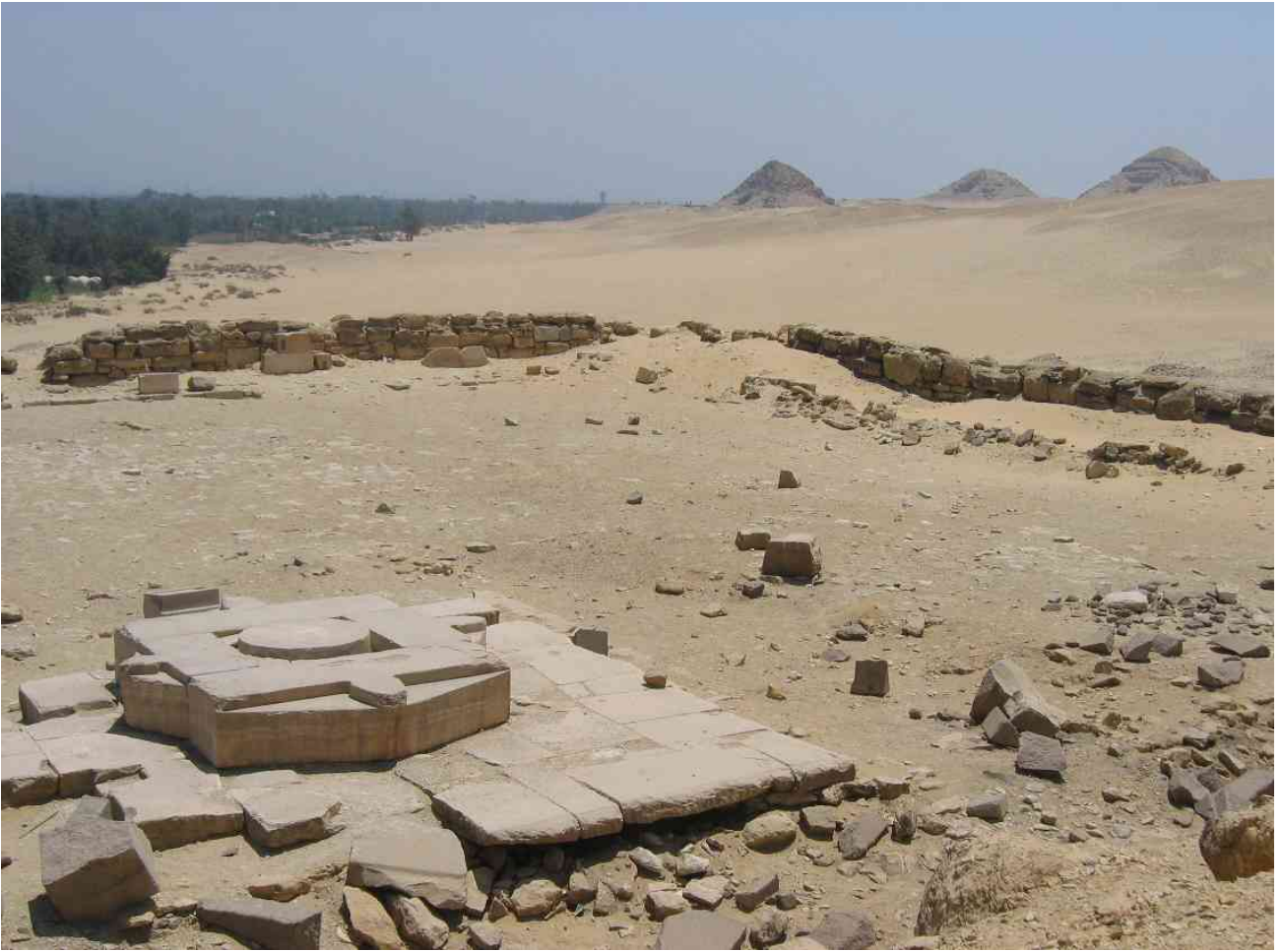

**Figure 11**
**The alabaster altar of the sun Temple of Niuserra, with the Abusir pyramids at the horizon. (Photograph by the author)**

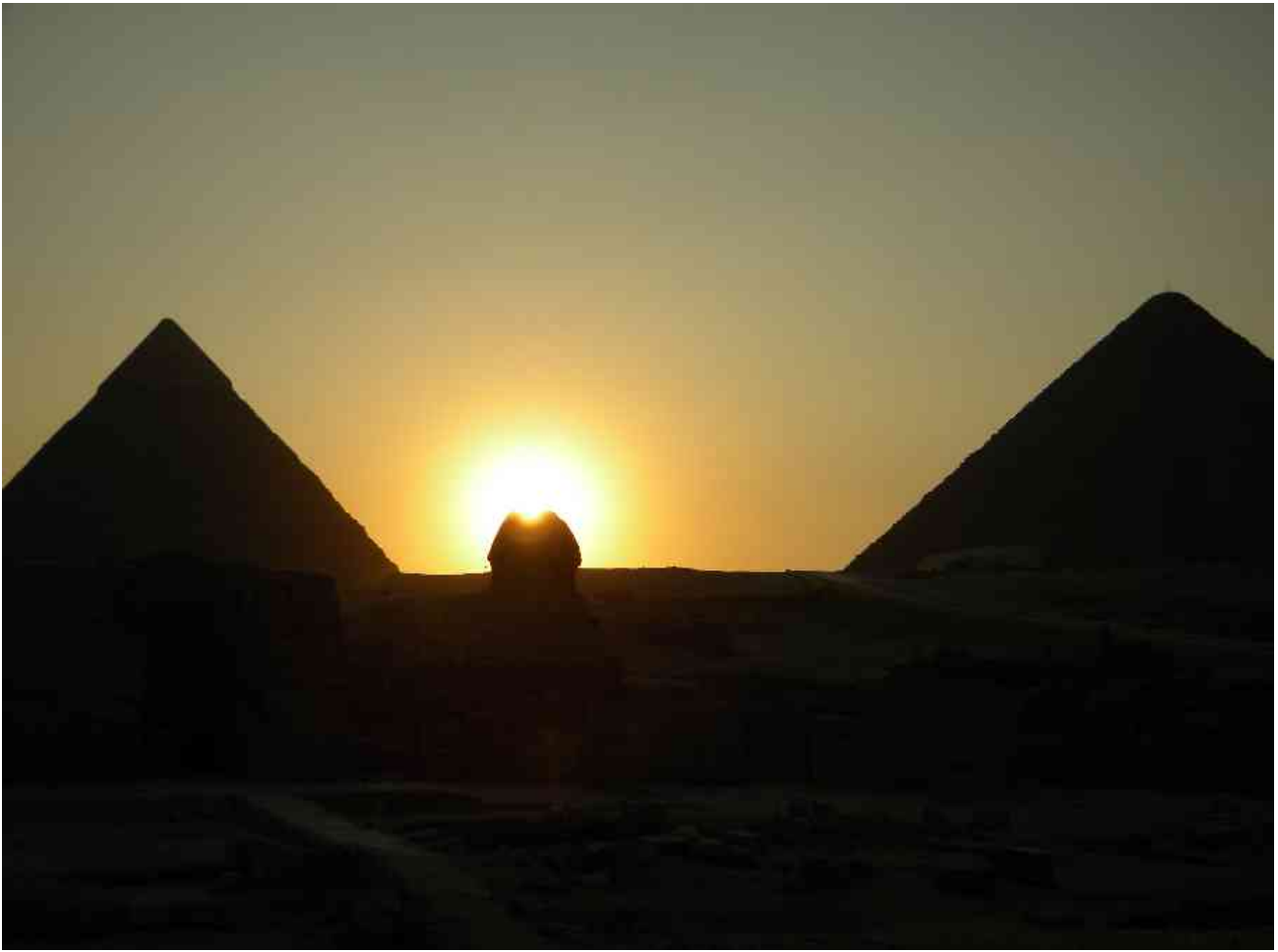

**Figure 12**
Summer sunset at Giza, viewed from the terrace in front of the Sphinx. In these moments the huge monument really becomes *Hor-em-akhet,* Horus in the Akhet, the way it was called in the New Kingdom. Clearly, the Akhet in question is the giant "two mountains plus sun" image forming at the rear of the statue, which is more and more similar to the corresponding hieroglyph in the days close to the summer solstice. (Photograph by the author)